\documentclass[preprint,12pt,authoryear]{elsarticle}
\usepackage{framed,multirow}

\usepackage{amssymb}
\usepackage{amsmath}
\usepackage{url}
\usepackage{xcolor}

\usepackage{algorithm} 
\usepackage{algpseudocode}
\usepackage{graphicx}
\usepackage{textcomp}
\usepackage{tabularx}
\usepackage{multirow}
\usepackage{natbib}
\usepackage{hyperref}
\usepackage{apalike}
\usepackage{xpatch}

\makeatletter
\xpatchcmd\NAT@citex
 {%
  \@citea\NAT@hyper@{%
    \NAT@nmfmt{\NAT@nm}%
    \hyper@natlinkbreak{\NAT@aysep\NAT@spacechar}{\@citeb\@extra@b@citeb}%
    \NAT@date
  }%
 }
 {%
  \@citea
  \NAT@nmfmt{\NAT@nm}%
  \NAT@aysep\NAT@spacechar
  \NAT@hyper@{\NAT@date}%
 }
 {}{}
\xpatchcmd\NAT@citex
 {%
  \@citea\NAT@hyper@{%
    \NAT@nmfmt{\NAT@nm}%
    \hyper@natlinkbreak{\NAT@spacechar\NAT@@open\if*#1*\else#1\NAT@spacechar\fi}%
    {\@citeb\@extra@b@citeb}%
    \NAT@date
  }%
 }
 {
  \@citea
    \NAT@nmfmt{\NAT@nm}%
    \NAT@spacechar\NAT@@open\if*#1*\else#1\NAT@spacechar\fi
    \NAT@hyper@{\NAT@date}%
 }
 {}{}
\makeatother

\journal{Medical Image Analysis}

\begin{document}

\begin{frontmatter}

\title{DIOR-ViT: Differential Ordinal Learning Vision Transformer for Cancer Classification in Pathology Images}

\author{Ju Cheon Lee\textsuperscript{1}, Keunho Byeon\textsuperscript{1}, Boram Song\textsuperscript{2}, Kyungeun Kim\textsuperscript{2,3}, and Jin Tae Kwak\textsuperscript{1}}

\address{
\textsuperscript{1}School of Electrical Engineering, Korea University, Seoul, Republic of Korea \\
\textsuperscript{2}Department of Pathology, Kangbuk Samsung Hospital, Sungkyunkwan University School of Medicine, Seoul, Republic of Korea \\
\textsuperscript{3}Pathology Center, Seegene Medical Foundation, Seoul, Republic of Korea \\
}

\begin{abstract}
In computational pathology, cancer grading has been mainly studied as a categorical classification problem, which does not utilize the ordering nature of cancer grades such as the higher the grade is, the worse the cancer is. To incorporate the ordering relationship among cancer grades, we introduce a differential ordinal learning problem in which we define and learn the degree of difference in the categorical class labels between pairs of samples by using their differences in the feature space. To this end, we propose a transformer-based neural network that simultaneously conducts both categorical classification and differential ordinal classification for cancer grading. We also propose a tailored loss function for differential ordinal learning. Evaluating the proposed method on three different types of cancer datasets, we demonstrate that the adoption of differential ordinal learning can improve the accuracy and reliability of cancer grading, outperforming conventional cancer grading approaches. The proposed approach should be applicable to other diseases and problems as they involve ordinal relationship among class labels.
\end{abstract}


\end{frontmatter}

\section{Introduction}
\label{sec:introduction}

Cancer is one of the leading causes of death worldwide \citet{bray2021ever}. In 2020, there were 19.3 million new cases and 10 million deaths from cancer globally \citet{sung2021global}. Once suspicious for cancer, tissue specimens are obtained from biopsy or surgery, stained with hematoxylin and eosin (H\&E), and evaluated by pathologists using a microscope. Although there has been a great deal of technical advances in medicine, a manual histological assessment of such tissue specimens still forms a definitive diagnosis, which often serves as a basis for cancer treatment and patient management. Since the current practice of pathology is, by and large, manually conducted, it is known to be low-throughput and subject to substantial inter- and intra-observer variations \citet{sooriakumaran2005gleason}, limiting the speed and accuracy of cancer diagnosis and, in turn, decreasing the quality of medical service. Hence, there is an urgent demand for developing a high-throughput, objective, and reliable method for cancer diagnosis.

Computational pathology, which integrates artificial intelligence, whole-slide imaging (WSI), and clinical informatics, is an emerging discipline in both clinical and scientific communities \citet{cui2021artificial}. Computational pathology tools have shown great potential for improving and reshaping the current practice of pathology \citet{rakha2021current}. Many of such tools are built based upon deep learning algorithms, in particular convolutional neural networks (CNNs). CNN-based methods have been successfully applied to several applications, including tissue segmentation \citet{vu2019dense}, mitosis detection \citet{sohail2021multi}, treatment response prediction \citet{hildebrand2021artificial}, nuclei segmentation and classification \citet{graham2019hover} \citet{doan2022sonnet} , and cancer grading \citet{le2021joint} \citet{ vuong2021multi}. Vision Transformer (ViT), on the other hand, has recently gained much attention for its superior performance in computer vision tasks, including object detection \citet{carion2020end} \citet{zhu2020deformable}, semantic segmentation \citet{zheng2021rethinking}, and scene understanding \citet{zhou2018end}. It has been also applied to pathology image analysis; for instance, cancer subtyping \citet{chen2022scaling}, metastasis detection \citet{gul2022histopathological}, and survival analysis\citet{shen2022explainable}.

Each cancer grade has its own unique histological and morphological patterns. Tissue samples of the same grade share such patterns. The role of pathologists or computational pathology tools is to identify the unique patterns and assign a pertinent grade or class label. In this sense, cancer grades or class labels have been typically studied as distinctive, independent categories, i.e., as a categorical classification problem. However, this neglects the relationship among cancer grades. There is a natural ordering among different cancer grades \citet{le2021joint}; the more aggressive the tumor cells are, the higher the cancer grade is. To incorporate such ordering relationship, cancer grading can be considered as an ordinal classification or regression problem, in which cancer grades are predicted on an ordinal scale. For example, \citet{le2021joint} formulated cancer grading as both categorical classification problem and ordinal classification problem and conducted the two classification tasks simultaneously. On the other hand, such ordering relationship can be realized through pair-wise comparisons among tissue samples of the same and/or differing grades. This approach of learning the ordering relationship is called order learning \citet{lim2020order}. Ordering learning has shown to be effective in age estimation \citet{lim2020order}. However, to the best of our knowledge, order learning and ordinal classification have not been fully exploited for cancer grading in computational pathology.

Herein, we propose a \textbf{DI}fferential \textbf{OR}dinal classification learning framework for \textbf{Vi}sion \textbf{T}ransformer (DIOR-ViT) that can conduct cancer grading in pathology images in an accurate and reliable fashion. The proposed DIOR-ViT is built based upon the recent development of the network architecture, i.e., ViT, and two learning paradigms, including multi-task learning and order learning. In DIOR-ViT, ViT is utilized to map an input tissue sample into a high-dimensional feature space for an efficient and effective representation of the sample. Under the principle of multi-task learning, DIOR-ViT simultaneously performs categorical cancer classification and differential ordinal cancer classification. 
\textcolor{black}{
In the categorical classification, DIOR-ViT predicts a class label for a given input sample. Meanwhile, in the differential ordinal classification, it receives an additional sample, designated as a subtrahend sample, and predicts the difference in the class labels between the two samples. 
}
For the categorical cancer classification, DIOR-ViT aims at capturing specific histological patterns of the tissue sample and determining its pertinent class, i.e., cancer grade. In the differential ordinal cancer classification, DIOR-ViT learns the ordering relationship between tissue samples via the differential pair-wise comparisons, i.e., difference between two samples in the feature space and ground truth labels. 
For efficient and effective optimization of DIOR-ViT, we introduce a new loss function, called negative absolute difference log-likelihood (NAD) loss, that is tailored to the differential ordinal classification. In order to evaluate the performance of DIOR-ViT, we employed a variety of multi-tissue cancer datasets, including colorectal, prostate, and gastric tissues, and compared it with several state-of-the-art models in computer vision and computational pathology. The proposed DIOR-ViT not only demonstrates accurate and robust classification results on multiple cancer datasets and but also substantially outperforms all the competing models. 
The main contributions of our work are summarized as follows:
\begin{itemize}
    \item We propose a vision transformer-based differential ordinal classification learning framework for improved cancer grading in computational pathology. 
    \item We introduce differential ordinal classification learning where we define the differential relationship between tissue samples in the feature space and ground truth labels, learn the differential relationship via pair-wise comparisons, and conduct ordinal classification to aid in improving cancer grading.
    \item We propose negative absolute difference log-likelihood loss that is tailored to the differential ordinal classification, leading to improved optimization of the proposed model. 
    \item We introduce a large gastric cancer dataset that includes $>$ 100,000 tissue images with four histopathologic classes such as benign, tubular well-differentiated tumor, tubular moderately-differentiated tumor, and tubular poorly-differentiated tumor.
    \item We assess the performance of the proposed method on three types of cancer datasets. The proposed method achieves the superior classification performance to other competing models on all three types of cancer datasets.
\end{itemize}

\section{Related Work}
\label{sec:relatedwork}

\subsection{Cancer Diagnosis in Computational Pathology}
Cancer diagnosis has been extensively studied in computational pathology. Earlier works have focused on extracting hand-crafted features that can capture the shape, arrangement, and distribution of histological objects, including cells, nuclei, glands, and etc. \citet{madabhushi2016image}. Several machine learning algorithms such as decision tree/random forest \citet{doyle2012cascaded}, SVM \citet{tabesh2007multifeature}, and boosting \citet{kwak2017multiview} have been employed to utilize these hand-crafted features and to make an automated cancer diagnosis. Later, various CNN-based methods have been widely explored for cancer diagnosis due to its greater learning ability with no or minimal human intervention \citet{deng2020deep}. For example, a single-stream CNN has been employed for breast cancer detection \citet{cruz2014automatic}, prostate cancer detection \citet{kwak2017nuclear} and grading \citet{arvaniti2018automated}, and colon cancer sub-typing \citet{hamida2021deep}. Multi-scale CNNs have been also utilized for several applications; for instance, \citet{kosaraju2020deep} utilized multi-scale receptive fields from low and high magnification images to conduct gastric cancer grading; \citet{vuong2021multi} introduced a multi-scale binary pattern encoding scheme to combine multi-scale information and to conduct colorectal and prostate cancer grading. Moreover, multi-task learning, which simultaneously performs multiple related tasks, was incorporated into CNNs for cancer diagnosis using pathology images \citet{zhao2019weakly} \citet{le2021joint}. Graph CNNs have been introduced to exploit a graph representation of pathology images and to conduct cancer classification for colorectal cancer \citet{zhou2019cgc} and breast cancer \citet{pati2022hierarchical}. Recently, ViT has shown to be effective in processing and analyzing both natural images and pathology images.

\subsection{Vision Transformer (ViT)}
Transformer \citet{vaswani2017attention} is a type of deep neural network that is built based upon self-attention mechanism and is first applied to natural language processing tasks such as machine translation \citet{vaswani2017attention}, question answering \citet{guan2022block}, and text classification \citet{zhang2021fast}. Inspired by transformer, ViT \citet{dosovitskiy2020transformers} was proposed for image recognition and achieved the state-of-the-art performance. In addition to image recognition, it has been applied to other computer vision tasks, including object detection \citet{carion2020end} \citet{zhu2020deformable}, semantic segmentation \citet{zheng2021rethinking}, and scene understanding \citet{zhou2018end}. ViT has been also applied to computational pathology. For example, \citet{chen2022scaling} proposed hierarchical image pyramid transformer (HIPT) architecture to aggregate cell-level, patch-level, and region-level visual tokens and to conduct cancer subtyping. \citet{gul2022histopathological} combined ViT with self-supervised learning and multiple instance learning strategies to detect cancer metastasis in pathology images where self-supervised learning was utilized to improve feature representations of input patches and multiple instance learning was used to aggregate patch-level representations. \citet{liao2022swin} integrated Swin Transformer, a type of Transformer using shifted window-based attention mechanism, and prior attention network into Swin-PANet, which was applied to medical image segmentation such as gland segmentation and nuclei segmentation in pathology images and skin lesion segmentation.

\subsection{Order Learning}
Order learning aims to predict the order or rank of a sample via pair-wise comparisons \citet{lim2020order}. It was first applied to age estimation from facial images where a facial image is compared to a reference facial image to determine whether it is younger than, similar to, and older than the reference image. \citet{lee2021deep} extended it by decomposing the information of a sample into order-related and (non-order) identity-related features and by incorporating deep repulsive clustering to group samples with similar identity into clusters. \citet{lee2022geometric} combined order learning with metric learning to construct the embedding space that represents both direction and distance between samples. Though successful in rank estimation, order learning represents the relationship between samples by greater than, similar to, and less than only, which ignores the magnitude of the difference between samples. In DIOR-ViT, we incorporate the magnitude of the difference between samples into cancer classification by introducing so called differential ordinal learning.

\subsection{Ordinal Classification}
The objective of ordinal classification or regression is to predict a rank or a class label of a sample on an ordinal scale. Many of ordinal classification works have converted the ordinal classification problem into a series of binary classification problems. For example, \citet{niu2016ordinal} utilized a CNN that conducts multiple binary classifications for age estimation. Similarly, \citet{fu2018deep} conducted monocular depth estimation using an end-to-end CNN model. Some others cast it as a regression problem. For instance, \citet{de2020deep} proposed a soft label ordinal regression for prostate cancer grading in magnetic resonance imaging. \citet{li2021learning} proposed a probabilistic ordinal embedding to model the uncertainty of regression and to conduct age estimation, image aesthetics assessment, and historical image dating. In \citet{le2021joint}, cancer grading was performed as an ordinal classification task in addition to a categorical classification task. Similar to this, DIOR-ViT simultaneously conducts both categorical and differential ordinal classifications; however, the differential ordinal classification of DIOR-ViT aims at predicting the differential ordering relationship between tissue samples, not directly predicting the cancer grade. Hence, the differential ordinal classification of DIOR-ViT can be understood as a combination of order learning and ordinal classification. To the best of our knowledge, this is the first attempt to combine the two approaches in computational pathology.

\section{Methods}

    \begin{figure*}[h]
        \centering
        \includegraphics[width=1.0\textwidth]{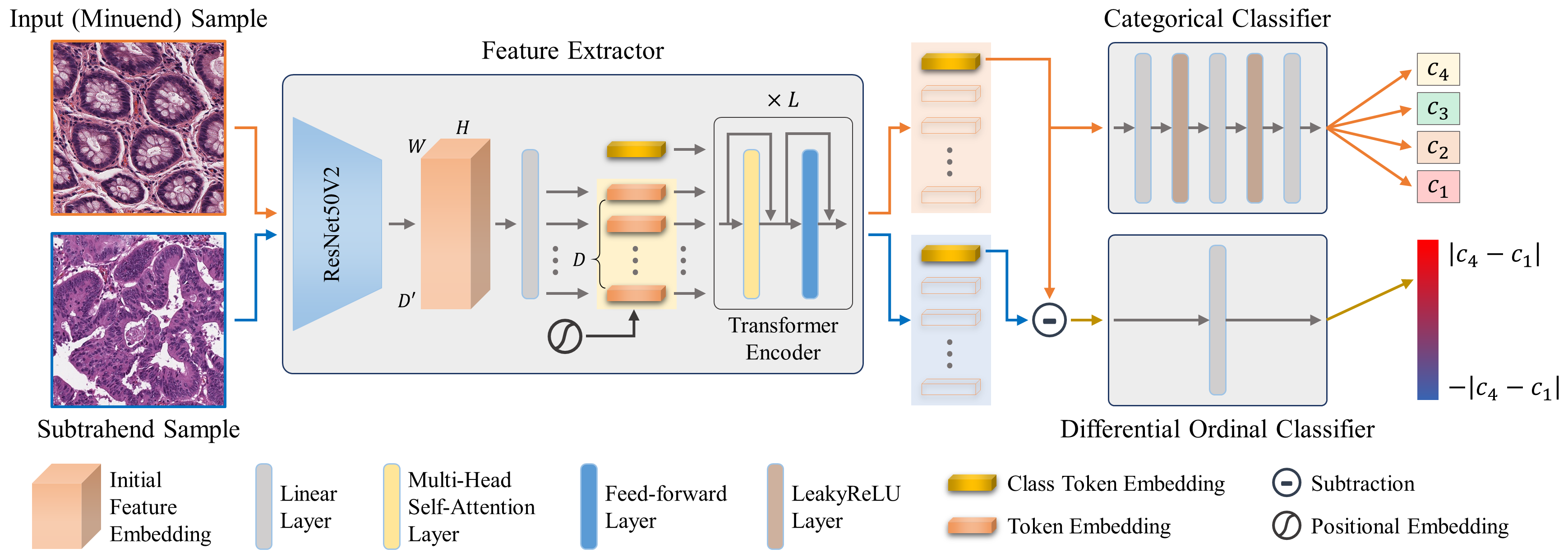}
        \caption{The overview of DIOR-ViT. DIOR-ViT consists of a feature extractor, a categorical classifier, and a differential ordinal classifier.}
        \label{fig:mod}
    \end{figure*}
    
\subsection{Problem Formulation}
Let \(\{{x}_{i}, {y}_{i}\}_{i=1}^{N}\) be a set of \(N\) pathology image-ground truth pairs where \({x}_{i}\) \(\in\) \({R}^{3}\) is the \(i\)th pathology image and \({y}_{i}\) \(\in\) \(\mathcal{C} = \{{c}_{1}, {c}_{2},…,{c}_{{n}_{c}}\}\) is the corresponding ground truth class label \({c}_{1}, {c}_{2},…,{c}_{{N}_{c}}\) are defined on an ordinal scale where \({N}_{c}\) is the cardinality of class labels; for instance, \({c}_{1}=1, {c}_{2}=2,…,{c}_{{N}_{c}}\) = \({N}_{c}\) (\({N}_{c}\) \(\in\) \(\mathcal{N}\)). The objective our study is to learn a transformer-based classifier $\mathcal{T}$ that maps an input sample \({x}_{i}\) into a high-dimensional feature space \(\Omega\), in which the ordering relationship among input samples are retained, and conducts cancer classification that are both accurate and reliable. It can be formulated as follows:
\begin{equation}
\underset{\theta}{argmin}\sum^{N}_{i=1}\mathcal{L}(\mathcal{T}({x}_{i}), {y}_{i};\theta)
\end{equation}
where $\mathcal{L}$ is the loss function and \(\theta\) is a set of learnable parameters of $\mathcal{T}$.

\subsection{Differential Ordinal Learning Vision Transformer (DIOR-ViT)}
The transformer-based classifier $\mathcal{T}$ contains three components, including a feature extractor $\mathcal{T}^{feat}$, a categorical classifier $\mathcal{T}^{cat}$, and a differential ordinal classifier $\mathcal{T}^{diff}$. Given an input sample \(x\), the feature extractor $\mathcal{T}^{feat}$ produces a high-dimensional feature vector \(f\). 
\textcolor{black}{
The categorical classifier $\mathcal{T}^{cat}$ receives the feature vector \(f\) and predicts the class label \(y\) (i.e., cancer grade) of the input sample \(x\). The differential ordinal classifier $\mathcal{T}^{diff}$ receives two feature vectors, a minuend feature vector \({f}^{m}\) from a minuend sample \({x}^{m}\) and a subtrahend feature vector \({f}^{s}\) from a subtrahend sample \({x}^{s}\), and compares the two feature vectors to predict the difference in the ground truth labels between \({x}^{m}\) and \({x}^{s}\), i.e., \(y^m - y^s\). 
}
Here, we designate the input sample \(x\) as the minuend sample and choose another input sample other than \(x\) as the subtrahend sample.

\subsubsection{Feature Extractor}
We build $\mathcal{T}^{feat}$ using ViT. Given an input sample \(x\), $\mathcal{T}^{feat}$ adopts ResNet50V2 \citet{he2016deep} to attain the initial high-dimensional feature embedding \({g}\in\mathcal{R}^{W×H×{D'}} (W=H=24, {D'}=1024)\), which undergoes a linear projection layer, producing a series of token embeddings of length \({E}\) \(\{{t}_{i} | i=1,…,E\} \in \mathcal{R}^{D}\) \((E=576, D=768)\). Then, we incorporate a class token \({t}^{CLS} \in \mathcal{R}^{D}\) to the token embeddings, generating \({g}_{0} = [{t}_{0}, {t}_{1}, {t}_{2},..., {t}_{E}]\), \({t}_{0} = {t}^{CLS}\). To add locational information, we adopt sin-cos positional embedding where we add cos(\({i}/{10000}^{{2j}/{D}}\)) and sin(\({i}/{10000}^{{2j+1}/{D}}\)) to every odd (\(2j+1\)) and even (\(2j\)) dimensions of \(i\)th token embedding, \(i=0,1,…,E\). These \(E+1\) token embeddings \({g}_{0}\) are used as the input to the transformer encoder. The transformer encoder utilizes a series of multi-head self-attention (\(MHSA\)) layers and Feed-forward (\(FF\)) layers to produce the output feature vector \(f\) as follows:
    \begin{equation}
        {g'}_{l} = MHSA({LN}({g}_{l-1})) + {g}_{l-1},      l = 1, ..., {L}
    \end{equation}

    \begin{equation}
        {g}_{l} = FF({LN}({g}_{l})) + {g'}_{l},            l = 1, ..., {L}
    \end{equation}

    \begin{equation}
        f={LN}(g^{0}_{L})
    \end{equation}
where \(LN\) represents a layer normalization layer, \({g}_{L}^{0}\) denotes the output embedding that corresponds to the class token \({t}^{CLS}\) at the \(L\)th block, \(L\) is the number of transformer encoder blocks. We set \(L\) to 12. \(MHSA\) conducts multiple self-attentions (SAs) in parallel given by:
    \begin{equation}
        MHSA(g) = Linear([{SA}_{1}(g), {SA}_{2}(g), ..., {SA}_{M}(g)])
    \end{equation}

    \begin{equation}
    \begin{split}
        {SA}_{i}(g) = softmax({qk}^{T} / \sqrt{{D}_{M}})v \\
        \text{where}\quad{q,k,v} = Linear(g), i=1, ..., M
    \end{split}    
    \end{equation}
\newline
where \(q\), \(k\), and \(v \in \mathcal{R}^{{D}_{M}}\) are query, key, and value vectors, respectively, \(M\) is the number of parallel SAs, and \({D}_{M}={D}⁄{M}\) . \(FF\) consists of three linear layers in which the first two are followed by a ReLU layer.

\begin{figure*}[h]
    \centering
    \includegraphics[width=1.0\linewidth]{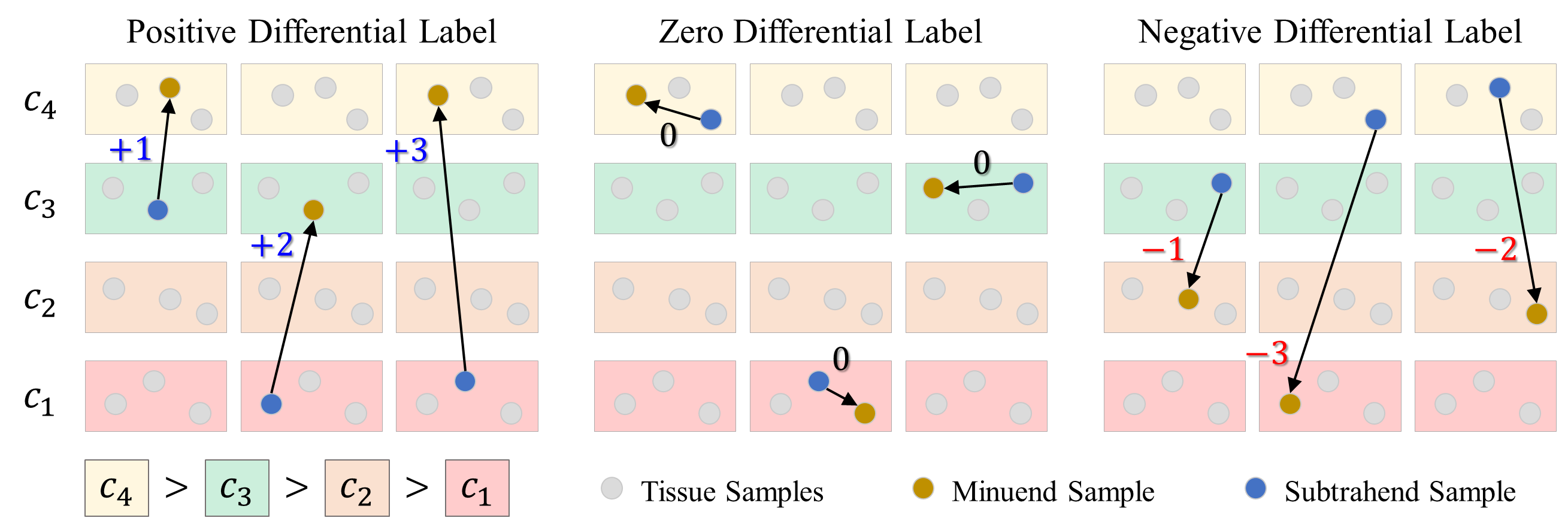}
    \caption{Generation of the ground truth differential ordinal labels.}
    \label{fig:ordinallabel}
\end{figure*}

\subsubsection{Categorical Classifier}
The categorical classifier $\mathcal{T}^{cat}$ consists of a linear layer, a LeakyReLU layer, a linear layer, a LeakyReLU layer, and a linear layer with \({N}_{c}\) neurons, followed by a softmax function. Provided with the feature vector \(f\), $\mathcal{T}^{cat}$ produces the output \({\hat{p}} = \{{\hat{p}}^{c} | c=1, ..., {N}_{c}\}\) where \(\hat{p}^{c}\) denotes the probability that the input sample belongs to the class \(c\).

\subsubsection{Differential Ordinal Classifier}
Let \({x}_{i}\) and \({x}_{j}\) be two input samples with class labels \({y}_{i}\) and \({y}_{j} \in {C}\), respectively. Following \citet{lim2020order}, the ordering relationship between \({x}_{i}\) and \({x}_{j}\) can be defined as:
    \begin{equation}
        \begin{split}
        {x}_{i} \succ   {x}_{j} \quad  \text{if} \quad {y}_{i} - {y}_{j} >  \tau \\
        {x}_{i} \approx {x}_{j} \quad  \text{if} \quad |{y}_{i} - {y}_{j}| \leq \tau \\
        {x}_{i} \prec   {x}_{j} \quad  \text{if} \quad {y}_{i} - {y}_{j} <    -\tau 
        \end{split} 
    \end{equation}
where \(\tau\) is a threshold value \(({\tau}=0.5)\) and \({x}_{i} \succ {x}_{j}\), \({x}_{i} \approx {x}_{j}\), and \({x}_{i} \prec {x}_{j}\) denote that the class label, i.e., cancer grade, of \({x}_{i}\) is higher than, equal to, and lower than that of \({x}_{j}\), respectively. The ordering relationship between samples can be informative in cancer classification since there is a natural ordering among cancer grades \citet{le2021joint}. However, these ordering relationships (\(\succ, \approx, \prec\)) ignore the degree of difference between the class labels. For instance, the difference between a high-grade cancer and a low-grade cancer should be considered to be much larger than the difference between a mid-grade cancer and a low-grade cancer. But, these two are considered to be equivalent in order learning. To incorporate the degree of difference between input samples, we define and utilize the differential relationship between the two samples using their categorical class labels as follows:
    \begin{equation}
        \gamma({x}_{i}, {x}_{j}): = {y}_{i} - {y}_{j}
    \end{equation}
where \(\gamma\) is a differential comparator, \({x}_{i}\) and \({y}_{i}\) denote the minuend input sample and its class label, respectively, and \({x}_{j}\) and \({y}_{j}\) represent the subtrahend input sample and its class label, respectively. Let \({r}_{i,j}\) be the ground truth differential relationship between the two input samples \({x}_{i}\) and \({x}_{j}\), i.e., \({r}_{i,j}={y}_{i}-{y}_{j}\). \({r}_{i,j}\) ranges from \(-|{c}_{{N}_{c}} - {c}_{1}|\) to \(|{c}_{{N}_{c}}-{c}_{1}|\). The magnitude of \({r}_{i,j}\) indicates how much the minuend input sample differs from the subtrahend sample. The sign of \({r}_{i,j}\) specifies the ordering relationship between the two samples. A positive, negative, and zero \({r}_{i,j}\) corresponds to  \(\succ, \prec, \approx\) in the ordering learning. Hence, the ordering relationships can be deduced from the differential relationships.

The objective of the differential classification is to approximate the differential comparator \(\gamma\) using the transformer-based classifier, i.e., $\mathcal{T}^{feat}$ and $\mathcal{T}^{diff}$. Given an input sample \({x}_{i}\), which is designated as a minuend sample, a feature vector \({f}^{i}\) is obtained by $\mathcal{T}^{feat}$. We choose another sample from a pool of input samples \(\{{x}_{j} | j=1,...,N, {j}\neq{i}\}\) , designated as a subtrahend sample, to be compared against the minuend sample. $\mathcal{T}^{feat}$ produces a feature vector \({f}_{j}\) for \({x}_{j}\). Provided with the two feature vectors \({f}_{i}\) and \({f}_{j}\), we subtract \({f}_{j}\) from \({f}_{i}\) and produce the differential feature vector \({f}_{{i},{j}}^{d} = {f}_{i} - {f}_{j}\). \({f}_{i,j}^{d}\) is fed into $\mathcal{T}^{diff}$ to predict the differential relationship between \({x}_{i}\) and \({x}_{j}\), i.e., \(\hat{r}_{i,j}\). For classification, $\mathcal{T}^{diff}$ contains a single linear layer with 1 neuron. 

During training, a batch of input samples \({X}^{b} = \{{x}_{i} |i=1,...,{N}_{b}\}\) and their categorical labels \({Y}^{b} = \{{y}_{i} |i=1,..,{N}_{b}\}\)
are provided with DIOR-ViT where \({N}_{b}\) is the size of the batch. For an input sample \({x}_{i}\), i.e., a minuend sample, each of all the other samples in \({X}^{b} \symbol{92} {x}_{i}\) is selected as a subtrahend sample one at a time and is used to compute the differential feature vector. Hence, we exploit the pair-wise differential relationships among input samples to learn the transformer-based classifier $\mathcal{T}$ that can re-arrange samples with respect to their degree of difference in the feature space, leading to improved categorical classification.

\subsection{Loss Function}
In order to optimize $\mathcal{T}$, we utilize two loss functions, one for each classification task, as follows:

\begin{equation}
\mathcal{L}_{total} = \mathcal{L}_{cat} + \lambda \mathcal{L}_{diff}
\end{equation}
where $\mathcal{L}_{cat}$ and $\mathcal{L}_{diff}$ denote the loss functions for the categorical classification and differential ordinal classification, respectively, and \(\lambda\) is a weight to balance the contribution of the two loss functions (\(\lambda = 6.5\)). For $\mathcal{L}_{cat}$, we adopt cross-entropy loss $\mathcal{L}_{CE}$ given by
\begin{equation}
\mathcal{L}_{CE}(p, \hat{p}) = -\frac{1}{N}\sum^{N}_{i=1}\sum^{N_c}_{c=1}p^c_i \texttt{log} \hat{p}^c_i
\end{equation}
where \({p}_{i}^{c} = 1\) if \({c}={y}_{i}\) otherwise \({p}_{i}^{c} = 0\) and \(\hat{p}_{i}^{c}\) represents the probability that \({x}_{i}\) belongs to the class \(c\).

\textcolor{black}{
As for $\mathcal{L}_{diff}$, though conventional regression loss functions such as a mean absolute error (MAE) and a mean square error (MSE) can be utilized, it has been shown that such functions are insufficient for the ordinal classification, in particular, when these are used with the categorical classification \citet{le2021joint}. 
An ordinal cross entropy loss function has been proposed to provide a clearer distinction between correct and wrong classifications and shown to be effective in improving the learning capability of the ordinal classification \citet{le2021joint}; however, this requires two steps: the conversion of the distance into probability measures by using a softmax function and calculation of the cross entropy loss for the probability measures; it is not smooth over the entire range of the distance. 
Inspired by \citet{le2021joint}, we propose a new loss function, negative absolute difference log-likelihood loss $\mathcal{L}_{NAD}$ that penalizes incorrect predictions on a logarithmic scale without extra steps. 
$\mathcal{L}_{NAD}$ is given as follows (Fig. 3):
\begin{equation}
\mathcal{L}_{NAD}(r, \hat{r}) = -\frac{1}{|\mathcal{P}|}\sum_{(i,j) \in \mathcal{P}}log(1 - \frac{|r_{i,j} - \hat{r}_{i,j}|} {2\mathcal{K} + \epsilon})
\end{equation}
where \(\mathcal{P}\) is a set of all possible input pairs, \({r}_{i,j}\) and \(\hat{r}_{i,j}\) denote the ground truth differential label and the predicted differential label for the input pair \({x}_{i}\) and \({x}_{j}\), respectively, \(\mathcal{K} = |c_{{N_c}} - c_1|\), and \(\epsilon = 1e^-5\).
}

The training procedure of $\mathcal{T}$ is provided in Algorithm 1.

\begin{figure}[t!]
    \centering
    \includegraphics[width=0.8\linewidth]{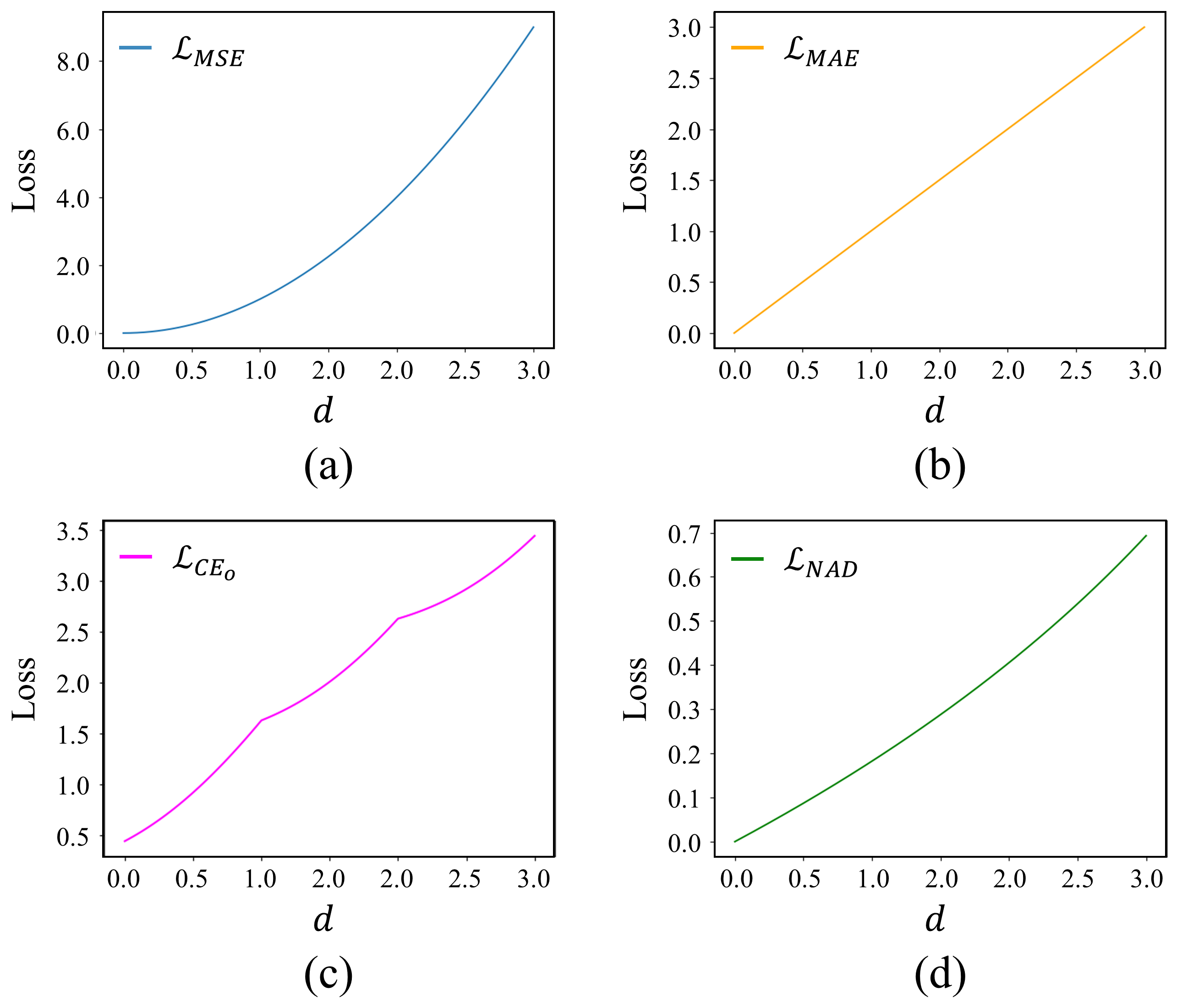}
    \caption{Loss curves for (a) mean squared error $\mathcal{L}_{MSE}$, (b) mean absolute error $\mathcal{L}_{MAE}$, (c) ordinal cross entropy loss $\mathcal{L}_{CE_o}$, and (d) negative absolute difference log-likelihood loss $\mathcal{L}_{NAD}$. $d$ denotes the difference between the ground truth label and the predicted label in the differential ordinal classification. }
    \label{fig:loss}
\end{figure}

\begin{algorithm}
\caption{Training procedure for DIOR-ViT}
\textbf{Input}: A batch of images \({X}^{b}=\{{x}_{i} |i=1,...,{N}_{b}\}\)  and their categorical labels \({Y}^{b}=\{{y}_{i} |i=1,...,{N}_{b}\}\), Feature extractor $\mathcal{T}^{feat}$, Categorical classifier $\mathcal{T}^{cat}$, Differential ordinal classifier $\mathcal{T}^{diff}$, Weight \(\lambda\) \\
\textbf{Output}: Updated $\mathcal{T}^{feat}$, $\mathcal{T}^{cat}$, and $\mathcal{T}^{diff}$

\label{alg:example}
\begin{algorithmic}[1]
    \For {all ${x}_{i}$ $\in$ ${X}^{b}$ and ${y}_{i}$ $\in$ ${Y}^{b}$}
        \State ${f}_{i}$ $\leftarrow$ $\mathcal{T}^{feat}({x}_{i})$ $\in$ $\mathbb{R}^{D}$
		\State $\hat{p}_{i}$ $\leftarrow$ $\mathcal{T}^{cat}({f}_{i})$ 
		\For {all ${x}_{j}$ $\in$ ${X}^{b}$ $\setminus$ ${x}_{i}$ and ${y}_{j}$ $\in$ ${Y}^{b}$ $\setminus$ ${y}_{i}$}
		\State ${r}_{i,j}$ $\leftarrow$ ${y}_{i}$ - ${y}_{j}$
		\State ${f}_{i}$ $\leftarrow$ $\mathcal{T}^{feat}({x}_{i})$ $\in$ $\mathbb{R}^{D}$
		\State ${f}_{i,j}^d$ $\leftarrow$ ${f}_{i}$ - ${f}_{j}$ $\in$ $\mathbb{R}^{D}$
  	\State $\hat{r}_{i,j}$ $\leftarrow$ $\mathcal{T}^{diff}({f}_{i,j}^{d})$ 
        \EndFor
    \EndFor
    \State $\mathcal{L}_{\text{cat}}$ $\leftarrow$ $\mathcal{L}_{\text{CE}}(y_i, \hat{p}_i)$
    \State $\mathcal{L}_{\text{diff}}$ $\leftarrow$ $\mathcal{L}_{\text{NAD}}(r, \hat{r})$
    \State $\mathcal{L}_{\text{total}}$ $\leftarrow$ $\mathcal{L}_{\text{cat}}$ + $\lambda \mathcal{L}_{\text{diff}}$
    \State Update $\mathcal{T}^{feat}$, $\mathcal{T}^{cat}$, and $\mathcal{T}^{diff}$ to minimize $\mathcal{L}_{\text{total}}$ \\
    \textbf{Return}: $\mathcal{T}^{feat}$, $\mathcal{T}^{cat}$, and $\mathcal{T}^{diff}$

\end{algorithmic}
\end{algorithm}

\section{Experiments}

\subsection{Dataset}
We employ a number of datasets from three types of tissues, including colorectal, prostate, and gastric tissues, to evaluate the performance of the proposed DIOR-ViT. Table \ref{table:data} shows the details of the datasets.
\subsubsection{Colorectal Tissue Dataset}
A set of publicly available colorectal tissue images are adopted by \citet{le2021joint} (2021). This contains a training dataset, a validation dataset, and two test datasets \(({C}_{TestI}\) and \({C}_{TestII})\). The training dataset, validation dataset, and \({C}_{TestI}\) were obtained from 6 tissue microarrays (TMAs) and 3 WSIs that were collected between 2006 and 2008. These were digitized at 40x magnification using an Aperio digital slide scanner (Leica Biosystems) with a pixel resolution of 0.2465 µm \(\times\) 0.2465 µm and 0.2518 µm \(\times\) 0.2518 µm for TMAs and WSIs, respectively. The training dataset, validation dataset, and \({C}_{TestI}\) consists of 7,027, 1,242, and 1,588 tissue images of size 1024 \(\times\) 1024 pixels ($\sim$258 µm \(\times\) 258 µm), respectively. \({C}_{TestII}\)  was acquired from 45 WSIs between 2016 and 2017 and was digitized at 40x magnification using a NanoZoomer digital slide scanner (Hamamatsu Photonics K.K.) with a pixel resolution of 0.2253 µm \(\times\) 0.2253 µm. \({C}_{TestII}\) comprises 110,171 tissue images of size 1024 \(\times\) 1024 pixels ($\sim$258 µm \(\times\) 258 µm). Each of the tissue images was assigned one of four class labels such as benign (BN), well-differentiated (WD) tumor, moderately-differentiated (MD) tumor, and poorly-differentiated (PD) tumor by pathologic review. 

\subsubsection{Prostate Tissue Dataset}
We utilize two public prostate tissue datasets. The first dataset is available at Harvard dataverse (https://dataverse.harvard.edu/), which was initially collected from 5 TMAs, digitized at 40x magnification (0.23 µm \(\times\) 0.23 µm per pixel) using a NanoZoomer digital slide scanner (Hamamatsu Photonics K.K.). 4 TMAs are used to construct a training dataset (2,107 tissue images) and a validation dataset (2,482 tissue images). The remaining TMA is used to build a test dataset (\({P}_{TestI}\); 4,237 tissue images). The size of each tissue image is 750 \(\times\) 750 pixels (172.5 µm \(\times\) 172.5 µm). The second dataset was obtained from the Gleason 2019 challenge (https://gleason2019.grand-challenge.org/). This is utilized as the second prostate test dataset (\({P}_{TestII}\)). \({P}_{TestII}\) is composed of 41,141 tissue images that were obtained from 244 tissue cores and digitized at 40x magnification (0.25 µm \(\times\) 0.25 µm per pixel) using an Aperio digital slide scanner (Leica Biosystems). Each tissue image has a size of 690 \(\times\) 690 pixels (172.5 µm \(\times\) 172.5 µm). Prostate tissue images are annotated with four class labels, including BN, grade 3 (G3) tumor, grade 4 (G4) tumor, and grade 5 (G5) tumor by pathologic review.  

\subsubsection{Gastric Tissue Dataset}
The gastric tissue dataset was collected from Kangbuk Samsung Hospital (IRB No. 2021-04-035). This includes 98 gastric WSIs from 98 patients, obtained from 2016 to 2020. WSIs were scanned at 40x magnification using an Aperio digital slide scanner (Leica Biosystems) and de-identified. The size of each WSI is $\sim$100,000 \(\times\) $\sim$80,000 pixels (0.2635 µm \(\times\) 0.2635 µm per pixel). 114,806 tissue images of size 1024 \(\times\) 1024 pixels are extracted from the 98 WSIs. Each of tissue image, of size 1024 \(\times\) 1024 pixels (270 µm \(\times\) 270 µm), is reviewed by experienced pathologists (B. Song and K. Kim) and is classified into BN, tubular well-differentiated (TW) tumor, tubular moderately-differentiated (TM), and tubular poorly- differentiated (TP) tumor. The entire tissue images are divided into a training dataset (83,638 tissue images), a validation dataset (15,381 tissue images), and a test data (\({G}_{TestI}\); 15,787 tissue images). 

\begin{table}[ht]
    \centering
    \caption{Details of colorectal, prostate, and gastric prostate tissue datasets.}
    \setlength{\tabcolsep}{2pt} 
    \begin{tabular}{cccccc}
        \hline
        Tissue Type & Class & Training & Validation & TestI & TestII \\
        \hline
        \begin{tabular}[c]{@{}c@{}}Colorectal Tissue\end{tabular} & BN & 773 & 374 & 453 & 27,896 \\
        & WD & 1,866 & 264 & 192 & 8,394 \\
        & MD & 2,997 & 370 & 738 & 61,985 \\
        & PD & 1,391 & 234 & 205 & 11,896 \\
        \hline
        \begin{tabular}[c]{@{}c@{}}Prostate Tissue\end{tabular} & BN & 2,096 & 666 & 127 & 16,454 \\
        & G3 & 6,303 & 923 & 1,602 & 2,057 \\
        & G4 & 2,541 & 573 & 2,121 & 2,060 \\
        & G5 & 2,383 & 320 & 387 & 20,570 \\
        \hline
        \begin{tabular}[c]{@{}c@{}}Gastric Tissue\end{tabular} & BN & 20,883 & 8,398 & 7,955 & - \\
        & WD & 14,251 & 2,239 & 1,795 & - \\
        & MD & 20,815 & 2,370 & 2,458 & - \\
        & PD & 27,689 & 2,374 & 3,579 & - \\
        \hline
    \end{tabular}
    \label{table:data}
\end{table}

\subsection{Comparative Experiments}
We compare the proposed DIOR-ViT to several models that were recently developed in computer vision and computational pathology. The competing models include 1) seven CNN-based models and 2) four transformer-based models. The seven CNN-based models are three plain CNN models (ResNet50 \citet{he2016deep}, DenseNet121 \citet{huang2017densely}, and EfficientNet-B0 \citet{tanm2019rethinking}, one multiscale CNN model (MSBP-Net \citet{vuong2021multi}, two multi-task CNN models ($\mathcal{M}_{MSE-CE_o}$ and  $\mathcal{M}_{MAE-CE_o}$ \citet{le2021joint}, \textcolor{black}{
and a CNN-based self-ensemble model (GLRGC \citet{xue2022robust})}. 
The four transformer-based models are ViT, Swin Transformer (Swin), DeiT III \citet{avidan2022computer}, \textcolor{black}{and StoHisNet \citet{fu2022stohisnet}}. All of these models utilize $\mathcal{L}_{CE}$ for optimization. In addition to \(\mathcal{L}_{CE}\), the two multi-task models, i.e., $\mathcal{M}_{MSE-CE_o}$ and $\mathcal{M}_{MAE-CE_o}$, adopt the conventional regression loss functions, including the mean squared error ($\mathcal{L}_{MSE}$) and mean absolute error ($\mathcal{L}_{MAE}$). The two models also utilize the ordinal cross entropy loss function ($\mathcal{L}_{CE_o}$) that converts the regression results into probability measures and computes cross entropy loss. 
\textcolor{black}{GLRGC uses three additional loss functions, including the consistency loss ($\mathcal{L}_{con}$), contrastive loss ($\mathcal{L}_{global}$), and global inter-sample relationship alignment loss ($\mathcal{L}_{global}$)}.

\subsection{Evaluation Metrics}
In order to evaluate the performance of the models under consideration, we employ three evaluation metrics: 1) accuracy (Acc): the percentage of the correctly classified samples over the all tissue samples; 2) macro F1-score (F1): the average per-class F1-score where per-class F1-score is a harmonic mean of precision and recall; 3) quadratic weighted kappa (\({k}_{w}\)): a chance-adjusted measure of the agreement between the predictions and the ground truth labels.

\subsection{Implementation Details}
We implement all models under consideration and conduct experiments using Pytorch platform using two RTX 3090 GPUs. 
\textcolor{black}{We utilize pretrained weights from the ImageNet dataset as initial weights for all the models.} 
The models are optimized using Adam optimizer \(({\beta}_{1}:0.9, {\beta}_{2}:0.999, \epsilon:1.0e^-8)\) and Cosine annealing warm restart scheduler (learning rate: \(1.0e^-4\)) for 50 epochs with a batch size of 16 batch size. Pre-trained weights on ImageNet dataset are adopted as the initial weights. All the tissue images are resized to 384 \(\times\) 384 pixels. Several data augmentation methods are applied using the Aleju library, including Affine transformation, one of Gaussian Blur, Average Blur, and Median Blur, Gaussian Noise, Dropout, linear contrast, horizontally and vertically flipping, and random saturation. 

\section{Results}

\subsection{Colorectal Cancer Classification}
Table \ref{table:colon} shows the classification results of the proposed DIOR-ViT and several other competing models on two independent colorectal cancer datasets (\({C}_{TestI}\) and \({C}_{TestII}\)). The results clearly demonstrate that the proposed DIOR-ViT were superior to other competing models. On \({C}_{TestI}\), DIOR-ViT achieved the best performance in Acc (87.78\%) and \({k}_{w}\)(0.942) and was the runner-up in F1 to one of the multi-task learning models $\mathcal{M}_{MAE-CE_o}$ by 0.002. Among the competing models, the two multi-task learning models outperformed others in all evaluation metrics. Moreover, on \({C}_{TestII}\), DIOR-ViT outperformed all the competing models by a large margin such as $\geq{5.26\%}$ Acc $\geq{0.030}$ F1, and $\geq{0.027}$ \({k}_{w}\) (Fig. \ref{fig:compare}a). It is noteworthy that \({C}_{TestI}\) and \({C}_{TestII}\) were acquired from different time periods and using different digital slide scanners. The superior performance of DIOR-ViT on \({C}_{TestII}\) indicates the better generalizability of the model under different acquisition settings. Unlike the results on \({C}_{TestI}\), the two transformer-based models (ViT and Swin) demonstrated the best performance among other competing models; however, another transformer-based model (DeiT III) was the worst model.

\begin{table}[ht]
\centering
\caption{Result of colorectal cancer classification.}
\setlength{\tabcolsep}{2pt} 
\begin{tabular}{ccccccc}
\hline
{Model}& \multicolumn{3}{c}{ \({C}_{TestI}\)}                       & \multicolumn{3}{c}{ \({C}_{TestII}\)}                       \\ \cline{2-7}     
                       & Acc(\%)        & F1             & \({k}_{w}\)              & Acc(\%)        & F1             & \({k}_{w}\)              \\ \hline
ResNet                 & 85.58          & 0.818          & 0.928          & 73.80           & 0.702          & 0.851          \\
DenseNet               & 85.33          & 0.827          & 0.929          & 75.40           & 0.703          & 0.867          \\
EfficientNet           & 85.89          & 0.815          & 0.924          & 73.50           & 0.694          & 0.845          \\
MSBP-Net               & 84.51          & 0.797          & 0.922          & 68.10           & 0.628          & 0.738          \\
\({M}_{MSE-{CE}_{o}}\)                & 87.30          & 0.832          & 0.941          & 74.90           & 0.694          & 0.849          \\
\({M}_{MAE-{CE}_{o}}\)                & 87.59          & \textbf{0.839} & 0.941          & 75.87          & 0.705          & 0.854          \\
ViT                    & 86.27          & 0.829          & 0.931          & 77.54          & 0.712          & 0.874          \\
Swin                   & 85.26          & 0.820          & 0.931          & 77.10          & 0.721          & 0.868          \\
Deit III               & 76.76          & 0.673          & 0.794          & 48.42          & 0.396          & 0.271          \\
\textcolor{black}{StoHisNet}              & 84.57          & 0.807          & 0.916          & 69.31          & 0.632          & 0.788          \\
\textcolor{black}{GLRGC}                  & 64.23          & 0.592          & 0.314          & 55.06          & 0.404          & 0.117          \\
DIOR-ViT(Ours)         & \textbf{87.78} & 0.837          & \textbf{0.942} & \textbf{82.80} & \textbf{0.751} & \textbf{0.901} \\
\hline
\end{tabular}
\label{table:colon}
\end{table}

\subsection{Prostates Cancer Classification}
\begin{figure*}[t!]
    \centering
    \includegraphics[width=\linewidth]{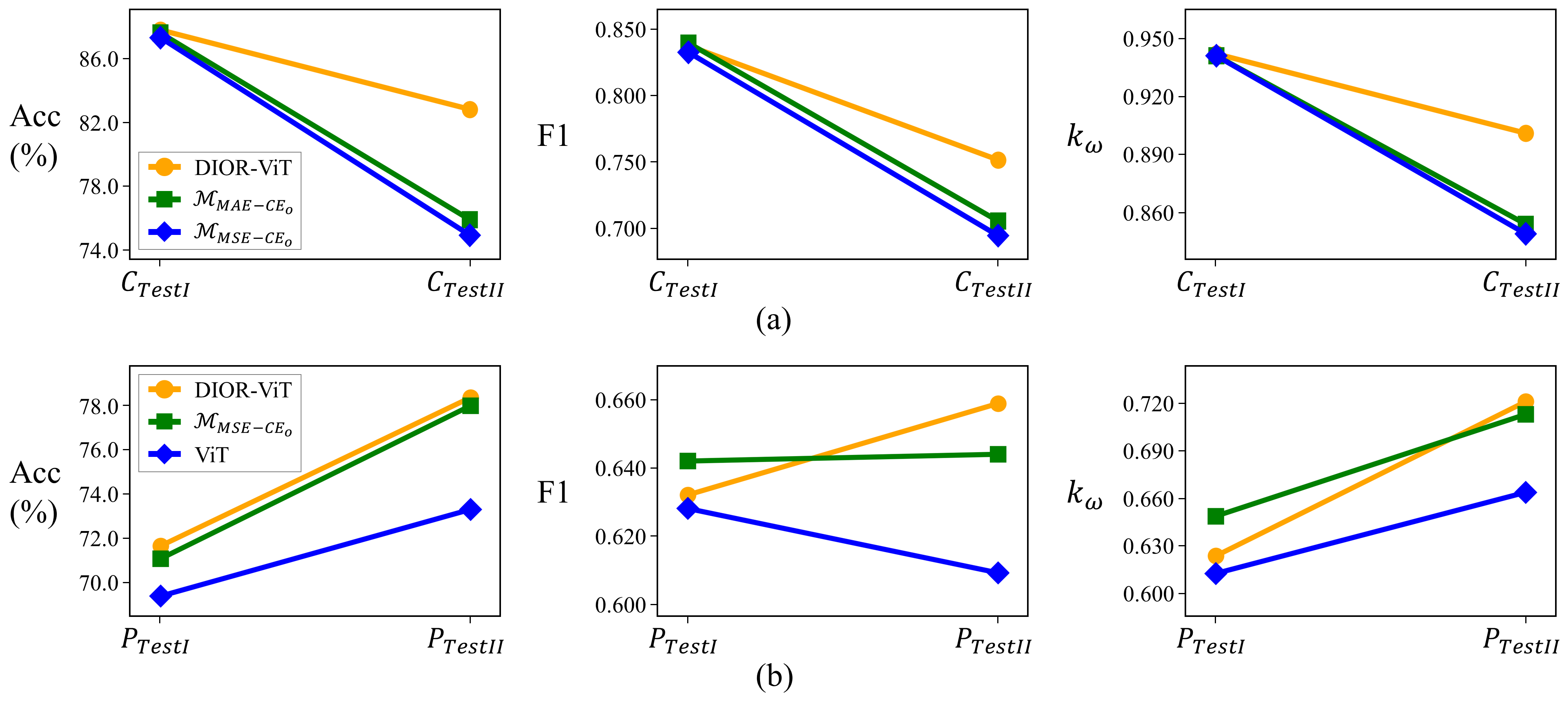}
    \caption{Comparison between Top-3 models on (a) colorectal and (b) prostate tissue datasets. }
    \label{fig:compare}
\end{figure*}

Table \ref{table:prostate} depicts the prostate cancer classification results on \({P}_{TestI}\) and \({P}_{TestII}\) by DIOR-ViT and other competing models. On \({P}_{TestI}\), DIOR-ViT obtained the best Acc of 71.64\%; however, $\mathcal{M}_{MAE-CE_o}$ and $\mathcal{M}_{MSE-CE_o}$ attained the best F1 of 0.644 and \({k}_{w}\) of 0.649, respectively. Except the two multi-task learning models, none of the competing models was comparable to DIOR-ViT. On \({P}_{TestII}\), we made similar observations with the results of colorectal cancer classification (Fig. \ref{fig:compare}b). DIOR-ViT outperformed all the competing models regardless of the evaluation metrics ($\geq{0.37}\%$ Acc, $\geq{0.015}\%$ F1, and $\geq{0.008}\%$ \({k}_{w}\)). This is striking because \({P}_{TestI}\) and \({P}_{TestII}\) were collected from different institutes and scanned using different slide scanners. These results also confirm the robustness of DIOR-ViT across different source of datasets. However, the performance of the competing models was generally inconsistent and unsteady. Though $\mathcal{M}_{MSE-CE_o}$ was still the second best model on \({P}_{TestI}\), $\mathcal{M}_{MAE-CE_o}$ was inferior to ViT. The two transformer-based models (ViT and Swin) were comparable to other competing models but DeiT III was again the worst model on both \({P}_{TestI}\) and \({P}_{TestII}\)

\begin{table}[ht]
\centering
\caption{Result of prostate cancer classification.}
\setlength{\tabcolsep}{2pt} 
\begin{tabular}{ccccccc}
\hline
{Model} & \multicolumn{3}{c}{\({P}_{TestI}\)}                       & \multicolumn{3}{c}{\({P}_{TestII}\)}                       \\ \cline{2-7} 
                       & Acc(\%)        & F1             & \({k}_{w}\)             & Acc(\%)        & F1             & \({k}_{w}\)             \\ \hline
ResNet                 & 66.09          & 0.582          & 0.561          & 71.55          & 0.562          & 0.621          \\
DenseNet               & 65.60          & 0.589          & 0.564          & 70.43          & 0.574          & 0.631          \\
EfficientNet           & 66.42          & 0.582          & 0.578          & 66.07          & 0.556          & 0.608          \\
MSBP-Net               & 62.26          & 0.532          & 0.491          & 61.61          & 0.441          & 0.446          \\
$\mathcal{M}_{MSE-CE_o}$                & 71.06          & 0.642          & \textbf{0.649} & 77.98          & 0.644          & 0.713          \\
$\mathcal{M}_{MAE-CE_o}$                & 69.37          & \textbf{0.664} & 0.646          & 72.54          & 0.632          & 0.639          \\
ViT                    & 69.37          & 0.628          & 0.613          & 73.29          & 0.609          & 0.664          \\
Swin                   & 66.45          & 0.573          & 0.554          & 68.96          & 0.574          & 0.636          \\
Deit III                 & 51.79          & 0.427          & 0.363          & 48.91          & 0.339          & 0.279          \\
\textcolor{black}{StoHisNet}              & 66.48          & 0.598          & 0.579          & 69.43          & 0.556          & 0.614          \\
\textcolor{black}{GLRGC}                  & 60.40          & 0.514          & 0.501          & 48.45          & 0.317          & 0.280          \\
DIOR-ViT(Ours)         & \textbf{71.64} & 0.632          & 0.624          & \textbf{78.35} & \textbf{0.659} & \textbf{0.721}    \\
\hline
\end{tabular}
\label{table:prostate}
\end{table}

\subsection{Gastric  Cancer Classification}
Table \ref{table:gastric} demonstrates the results of the gastric cancer classification by the models under consideration. Similar trends were found with the results on  \({C}_{TestII}\) and \({P}_{TestII}\). DIOR-ViT was superior to all the competing models in all evaluation metrics. Comparing to other competing models, the classification performance was improved by $\geq{0.47}\%$ Acc, $\geq{0.006}$ F1, and $\geq{0.003}$ \({k}_{w}\). Among the competing models, \({M}_{MSE-{CE}_{o}}\) achieved the second best F1 of 0.780 and \({k}_{w}\) of 0.933 and EfficientNet obtained the second best Acc of 85.01\%. But, the results were, by and large, comparable among all the competing models except DeiT III.

\begin{table}[]
\centering
\caption{Result of Gastric cancer classification.}
\centering
\begin{tabular}{cccc}
\hline
{Model} & \multicolumn{3}{c}{\({G}_{TestI}\)}                       \\ \cline{2-4} 
                       & Acc(\%)        & F1             & \({k}_{w}\)             \\ \hline
ResNet                 & 83.76          & 0.768          & 0.933          \\
DenseNet               & 84.52          & 0.778          & 0.932          \\
EfficientNet           & 85.01          & 0.774          & 0.914          \\
MSBP-Net               & 83.07          & 0.751          & 0.897          \\
$\mathcal{M}_{MSE-CE_o}$                & 84.51          & 0.780          & 0.933          \\
$\mathcal{M}_{MAE-CE_o}$                & 83.73          & 0.768          & 0.929          \\
ViT                    & 84.06          & 0.772          & 0.931          \\
Swin                   & 83.71          & 0.759          & 0.919          \\
Deit III                 & 77.05          & 0.656          & 0.847          \\
\textcolor{black}{StoHisNet}                 & 83.61          & 0.765          & 0.922          \\
\textcolor{black}{GLRGC}                 & 80.61          & 0.734          & 0.874          \\
DIOR-ViT(Ours)           & \textbf{85.48} & \textbf{0.786} & \textbf{0.936}   \\
\hline
\end{tabular}
\label{table:gastric}
\end{table}

\subsection{Analysis of DIOR-ViT}
\textcolor{black}{
In order to gain an in-depth understanding of the proposed DIOR-ViT, we conducted ablation experiments for the loss functions and feature extractors.} For the loss functions, we carried out three types of ablation experiments as follows: 1) Effect of the differential ordinal classification: DIOR-ViT without differential ordinal classification; 2) Effect of the regression loss function: DIOR-ViT that utilizes the conventional regression loss functions ($\mathcal{L}_{MSE}$ and $\mathcal{L}_{MAE}$) instead of $\mathcal{L}_{NAD}$; 3) Effect of the ordinal regression loss function: DIOR-ViT that utilizes conventional regression loss functions and ordinal cross entropy loss function ($\mathcal{L}_{{CE}_{o}}$) instead of $\mathcal{L}_{NAD}$. Table \ref{table:loss} provides the experimental results on colorectal, prostate, and gastric datasets. The first experiment eliminates the differential ordinal classification from DIOR-ViT, which, in fact, gives rise to ViT that is trained with $\mathcal{L}_{CE}$. As shown above, the contribution of the differential ordinal classification is obvious regardless of the datasets and evaluation metrics.
For the second type of the experiments, we employed two regression loss functions ($\mathcal{L}_{MSE}$ and $\mathcal{L}_{MAE}$) for the differential ordinal classifier. Equipped with $\mathcal{L}_{MSE}$ and $\mathcal{L}_{MAE}$, the performance of DIOR-ViT was substantially dropped by $\leq{4.20}\%$ Acc, $\leq{0.043}$ F1, and $\leq{0.043}$ \({k}_{w}\) on the colorectal tissue datasets, $\leq{2.03}\%$ Acc, $\leq{0.016}$ F1, and $\leq{0.024}$ \({k}_{w}\) on the prostate tissue datasets, and $\leq{2.15}\%$ Acc, $\leq{0.023}$ F1, and $\leq{0.017}$ \({k}_{w}\) on the gastric tissue dataset. 
As for the third type of the experiments, we exploited the ordinal cross entropy loss function $\mathcal{L}_{{CE}_{o}}$, which was proposed in [13]. Using $\mathcal{L}_{MSE}$ and $\mathcal{L}_{MAE}$ with $\mathcal{L}_{{CE}_{o}}$, the performance of the models was still inferior to DIOR-ViT with $\mathcal{L}_{NAD}$. For instance, on the colorectal and prostate tissue datasets, the classification performance decreased by $\leq{5.70}\%$ Acc, $\leq{0.026}$ F1, and $\leq{0.024}$ \({k}_{w}\) and by $\leq{5.76}\%$ Acc, $\leq{0.059}$ F1, and $\leq{0.053}$ \({k}_{w}\), respectively. On the gastric tissue dataset, Acc and F1 were dropped by at most 0.51\% and 0.002 F1, but \({k}_{w}\) was increased by 0.004 as the model utilizes $\mathcal{L}_{CE}+\mathcal{L}_{MSE}+\mathcal{L}_{CE_o}$. The second and third type of experimental results apparently demonstrate the utility of the proposed negative absolute difference log-likelihood loss $\mathcal{L}_{NAD}$ for the differential ordinal classification learning.

\textcolor{black}{
Moreover, we evaluated the effect of the feature extractor on DIOR-ViT. We replaced the feature extract of DIOR-ViT, i.e., ViT, by Swin, which is designated as DIOR-Swin. Using DIOR-Swin, we repeated the same experiments with DIOR-ViT. The results are available in Table \ref{table:backbone}. Across the three types of cancer datasets, DIOR-Swin was generally inferior to DIOR-ViT except for Acc and F1 in $C_{TestII}$. As we compared the results by ViT and Swin alone, we also found that Swin, in general, obtained lower performance than ViT, likely suggesting that the inferior performance of DIOR-Swin is partially due to the capability of Swin. These observations also indicate that the feature extractor plays a critical role in the proposed method. In a head-to-head comparison between Swin and DIOR-Swin and between ViT and DIOR-ViT, we found that the adoption of differential ordinal learning almost always increased the performance. For instance, in comparison to ViT, DIOR-ViT provided a substantial performance gain by $\leq$5.26\% Acc, $\leq$0.039 F1, and $\leq$0.027 \({k}_{w}\) on the colorectal tissue datasets, $\leq$5.06\% Acc, $\leq$0.050 F1, and $\leq$0.057 \({k}_{w}\) on the prostate tissue datasets, and 1.42\% Acc, 0.014 F1, and 0.005 \({k}_{w}\) on the gastric tissue dataset. Similar observations were made for the comparison between DIOR-Swin and Swin. 
}

\textcolor{black}{
To provide further insights into DIOR-ViT and the classification results, we visualized and analyzed the attention heatmaps of tissue images that were generated by GradCAM \citet{selvaraju2017grad}.
Figure \ref{fig:heatmap} shows some exemplary tissue images from \({C}_{TestII}\) and the corresponding attention heatmaps by ViT and DIOR-ViT. These attention heatmaps clearly show that DIOR-ViT focuses and utilizes pathologically important and relevant regions in decision-making. These are substantially different from those by ViT.
For instance, in the BN samples, both DIOR-ViT and ViT focus on the cytoplasm; however, DIOR-ViT demonstrates the uniform and strong attention, whereas ViT shows the weaker and disproportionate attention within glands. 
In the WD samples, in which tumor cells form glands well, DIOR-ViT clearly highlights most of tumor cells, but ViT shows scattered attentions around tumor cells including cytoplasm, stroma, and luminal areas. 
For the MD samples, ViT highlights scattered areas such as tumor and mucinous areas, meanwhile DIOR-ViT primarily focuses on the stroma areas that are denser or fibrotic due to tumor growth and invasion and contain infiltration of inflammatory cells. 
As for the PD samples, DIOR-ViT appears to focus on the entire samples, likely due to the fact that the samples are full of highly altered and irregularly arranged tumor cells with nuclei of significant atypia such as large, round, and vesicular nuclei and occasional distinctive nucleoli. ViT, on the other hand, partially highlights the samples, missing some of the tumor cells with clear patterns of PD.
}

\begin{figure*}[t!]
    \centering
    \includegraphics[width=\linewidth]{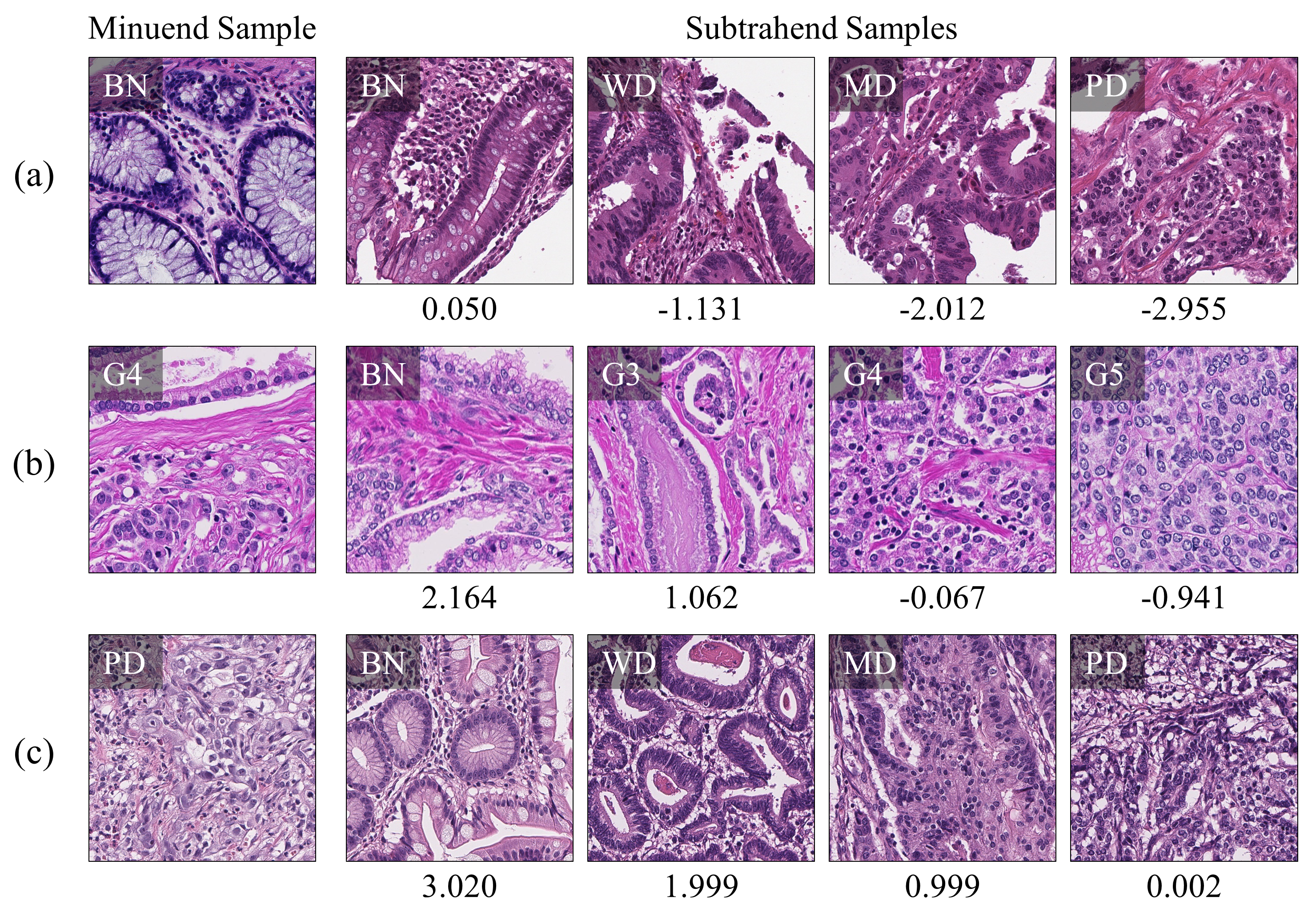}
    \caption{Differential ordinal classification results on (a) colorectal, (b) prostate, and (c) gastric tissue datasets. The number below each image represents the output of the differential ordinal classifier. }
    \label{fig:diff}
\end{figure*}

\begin{figure*}[t!]
    \centering
    \includegraphics[width=1.0\textwidth]{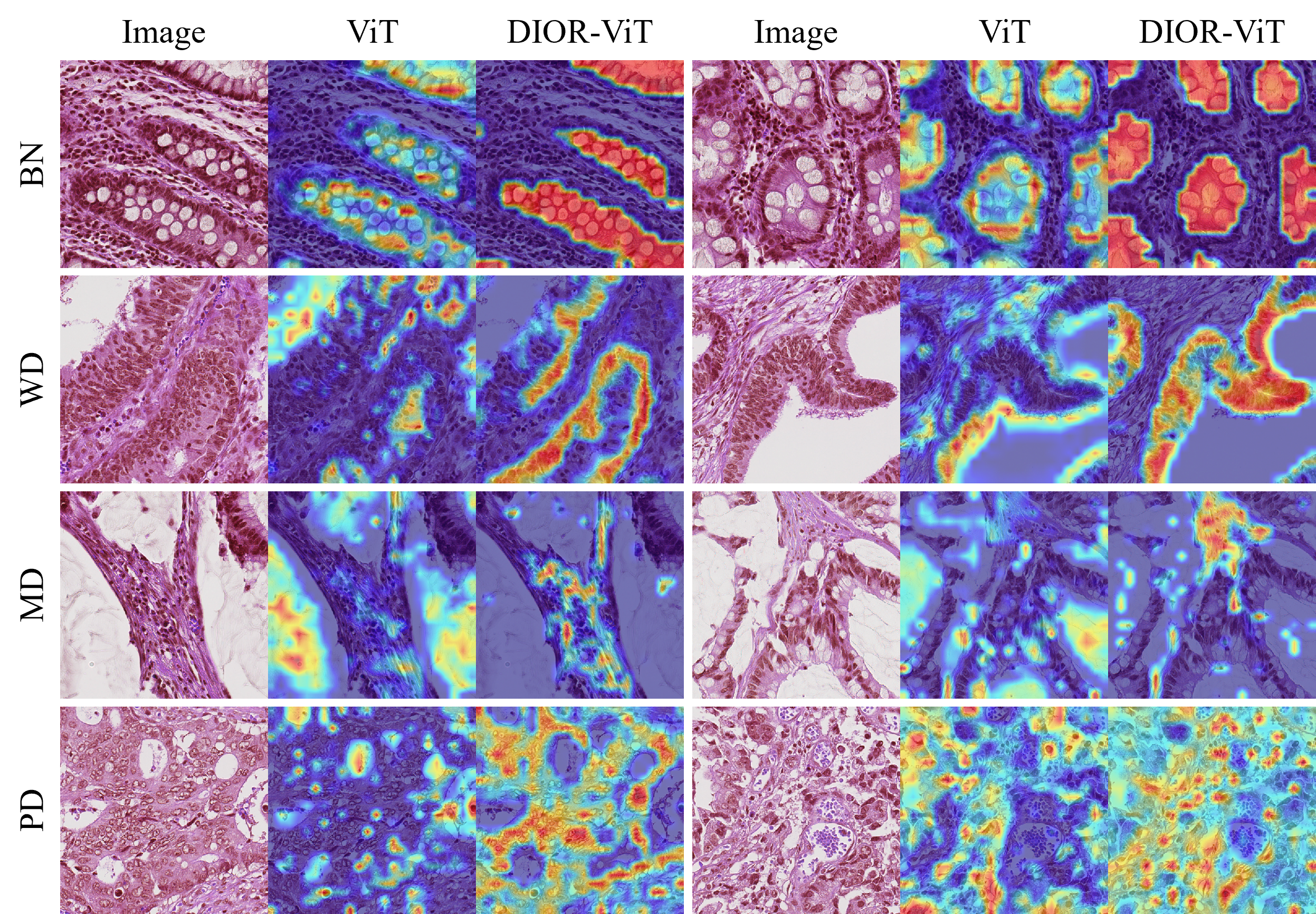}
    \caption{Example images and GradCAM results for each class from the \({C}_{TestII}\) dataset.  }
    \label{fig:heatmap}
\end{figure*}

\textcolor{black}{
Futhermore, we investigated the quality of the differential ordinal classification. Fig. \ref{fig:diff} depicts the exemplary minuend and subtrahend samples from the three cancer datasets with the output of the differential ordinal classifier. Each of the minuend samples was correctly classified into the pertinent class label by the categorical classifier of DIOR-ViT. As DIOR-ViT receives each pair of the minuend sample and the subtrahend sample, the differential ordinal classifier was able to predict the difference in the ground truth categorical class labels of the two samples with a small error. In other words, the differential ordinal classifier was capable of approximating the differential relationship between pairs of minuend and subtrahend samples.
}

\textcolor{black}{
In addition, we measured and compared the complexity of DIOR-ViT and other competitors using the number of parameters, floating-point operations per second (FLOPs) for both training and inference, and training and inference time per image. Table \ref{table:model_complexity} illustrates the quantitative measurements of the model complexity. 
In general, transformer-based models are shown to be larger than CNN-based models with respect to the number of parameters and FLOPs; since DIOR-ViT adopts ViT as a backbone network, it becomes one of the largest models. However, the training and inference time of DIOR-ViT are comparable to other competitors. It is noticeable that DIOR-ViT shows almost the same complexity with ViT for both training and inference. During training, DIOR-ViT separately and independently extracts embeddings vectors of the minuend and subtrahend samples, conducts a subtraction operation for each pair of the samples, and goes through a single linear layer. Though DIOR-ViT requires pair-wise comparisons among the input samples, the added complexity is minimal, suggesting the efficiency of the differential ordinal learning in DIOR-ViT.
}

\begin{table}[]
\centering
\caption{Result of cancer classification with differing loss functions.}
\setlength{\tabcolsep}{2pt} 
\begin{tabular}{clcccccc}
\hline
{Model}      & Loss & \multicolumn{3}{c}{\({C}_{TestI}\)}    & \multicolumn{3}{c}{\({C}_{TestII}\)}                   \\  \cline{3-8}
&            & Acc(\%)  & F1 & \({k}_{w}\)  & Acc(\%)        & F1             & \({k}_{w}\)           \\ \hline 
ViT          & $\mathcal{L}_{CE}$ &  86.27          & 0.829          & 0.931    & 77.54          & 0.712          & 0.874      \\
DIOR-ViT      & $\mathcal{L}_{CE} + \mathcal{L}_{MSE}$ &  86.90           & 0.832          & 0.937     & 81.20          & 0.744          & 0.893      \\
DIOR-ViT      & $\mathcal{L}_{CE} + \mathcal{L}_{MAE}$ &  85.14          & 0.794          & 0.931  & 78.60          & 0.715          & 0.878        \\
DIOR-ViT      & $\mathcal{L}_{CE} + \mathcal{L}_{MSE} + \mathcal{L}_{CE_o}$ &  86.52          & 0.834          & 0.939  & 77.10          & 0.725          & 0.877        \\
DIOR-ViT      & $\mathcal{L}_{CE} + \mathcal{L}_{MAE} + \mathcal{L}_{CE_o}$ &  87.15          & 0.836          & 0.936  & 79.90          & 0.745          & 0.893        \\
DIOR-ViT(Ours)& $\mathcal{L}_{CE} + \mathcal{L}_{NAD}$ &  \textbf{87.78} & \textbf{0.837} & \textbf{0.942}  & \textbf{82.80} & \textbf{0.751} & \textbf{0.901}  \\
\hline
{Model}      & Loss & \multicolumn{3}{c}{\({P}_{TestI}\)}    & \multicolumn{3}{c}{\({P}_{TestII}\)}                   \\  \cline{3-5} \cline{6-8}
&            & Acc(\%)  & F1 & \({k}_{w}\)   & Acc(\%)        & F1             & \({k}_{w}\)          \\ \hline 
ViT          & $\mathcal{L}_{CE}$ & 69.37          & 0.628          & 0.613    & 73.29          & 0.609          & 0.664      \\
DIOR-ViT      & $\mathcal{L}_{CE} + \mathcal{L}_{MSE}$ & 71.61       & 0.619       & 0.622  & 76.32          & 0.644       & 0.698         \\
DIOR-ViT      & $\mathcal{L}_{CE} + \mathcal{L}_{MAE}$ & 70.86       & 0.616       & 0.616  & 77.32          & 0.644       & 0.697        \\
DIOR-ViT      & $\mathcal{L}_{CE} + \mathcal{L}_{MSE} + \mathcal{L}_{CE_o}$ & 69.25    & 0.623    & 0.611 & 72.59      & 0.600  & 0.668  \\
DIOR-ViT      & $\mathcal{L}_{CE} + \mathcal{L}_{MAE} + \mathcal{L}_{CE_o}$ & 68.95    & 0.631    & 0.607 & 75.98      & 0.624  & 0.701   \\
DIOR-ViT(Ours)& $\mathcal{L}_{CE} + \mathcal{L}_{NAD}$ &  \textbf{71.64} & \textbf{0.632} & \textbf{0.624} & \textbf{78.35} & \textbf{0.659} & \textbf{0.721}  \\
\hline
{Model}                & {Loss} & \multicolumn{3}{c}{\({G}_{TestI}\)}                       \\ \cline{3-5} 
                       &                       & Acc(\%)        & F1             & \({k}_{w}\)             \\ \hline
ViT                    & $\mathcal{L}_{CE}$                    & 84.06          & 0.772          & 0.931          \\
DIOR-ViT                & $\mathcal{L}_{CE} + \mathcal{L}_{MSE}$  & 84.50           & 0.780         & 0.929     \\
DIOR-ViT                & $\mathcal{L}_{CE} + \mathcal{L}_{MAE}$  & 83.33          & 0.763          & 0.919      \\
DIOR-ViT                & $\mathcal{L}_{CE} + \mathcal{L}_{MSE} + \mathcal{L}_{CE_o}$  & 85.46     & 0.784          & \textbf{0.940} \\
DIOR-ViT                & $\mathcal{L}_{CE} + \mathcal{L}_{MAE} + \mathcal{L}_{CE_o}$  & 84.97     & 0.785    & 0.933   \\
DIOR-ViT(Ours)           & $\mathcal{L}_{CE} + \mathcal{L}_{NAD}$  & \textbf{85.48} & \textbf{0.786} & 0.936 \\
\hline
\end{tabular}
\label{table:loss}
\end{table}

\begin{table}[ht]
\centering
\caption{Result of cancer classification with differing feature extractors.}
\setlength{\tabcolsep}{2pt} 
\begin{tabular}{ccccccc}
\hline
{Model}& \multicolumn{3}{c}{ \({C}_{TestI}\)}                       & \multicolumn{3}{c}{ \({C}_{TestII}\)}                       \\ \cline{2-7}     
                       & Acc(\%)        & F1             & \({k}_{w}\)              & Acc(\%)        & F1             & \({k}_{w}\)              \\ \hline
ViT                    & 86.27          & 0.829          & 0.931          & 77.54          & 0.712          & 0.874          \\
Swin                   & 85.26          & 0.820          & 0.931          & 77.10          & 0.721          & 0.868          \\
DIOR-Swin(Ours)        & 87.09 & 0.825          & 0.938 & \textbf{83.44} & \textbf{0.755} & 0.900 \\
DIOR-ViT(Ours)         & \textbf{87.78} & \textbf{0.837} & \textbf{0.942} & 82.80 & 0.751 & \textbf{0.901} \\
\hline
{Model}& \multicolumn{3}{c}{ \({P}_{TestI}\)}                       & \multicolumn{3}{c}{ \({P}_{TestII}\)}                       \\ \cline{2-7}     
                       & Acc(\%)        & F1             & \({k}_{w}\)              & Acc(\%)        & F1             & \({k}_{w}\)              \\ \hline
ViT                    & 69.37          & 0.628          & 0.613          & 73.29          & 0.609          & 0.664          \\
Swin                   & 66.45          & 0.573          & 0.554          & 68.96          & 0.574          & 0.636          \\
DIOR-Swin(Ours)        & 69.11          & 0.630          & 0.605          & 72.86          & 0.574          & 0.713          \\
DIOR-ViT(Ours)         & \textbf{71.64} & \textbf{0.632} & \textbf{0.624}          & \textbf{78.35} & \textbf{0.659} & \textbf{0.721}    \\
\hline
{Model} & \multicolumn{3}{c}{\({G}_{TestI}\)}                       \\ \cline{2-4} 
                       & Acc(\%)        & F1             & \({k}_{w}\)             \\ \hline
ViT                    & 84.06          & 0.772          & 0.931          \\
Swin                   & 83.71          & 0.759          & 0.919          \\
DIOR-Swin(Ours)                 & 83.61          & 0.768          & 0.928          \\
DIOR-ViT(Ours)           & \textbf{85.48} & \textbf{0.786} & \textbf{0.936}   \\
\hline
\end{tabular}
\label{table:backbone}
\end{table}

\section{Discussion}
We introduce a differential ordinal learning problem for cancer classification in computational pathology. Exploiting the natural ordering of cancer grades, the proposed DIOR-ViT facilitates improved cancer classification in an accurate and robust manner.

Three types of cancer datasets were employed in this study. DIOR-ViT outperformed the 9 competing models that were built based upon different architectures and under different learning paradigms. Among other competing models, we observed that the results were inconsistent across datasets and cancer types. These observations emphasize the superiority and utility of the proposed DIOR-ViT in cancer grading and its applicability to other types of cancers and datasets in computational pathology. 

The colorectal and prostate datasets include the independent datasets ($C_{TestII}$ and $P_{TestII}$) that were collected from different acquisition settings and institutes. For both datasets, DIOR-ViT outperformed all other competing models, indicating the robustness of DIOR-ViT. However, there was a difference between the two datasets in terms of the model performance. On $C_{TestII}$, the performance of DIOR-ViT and other models substantially decreased in comparison to the results on $C_{TestI}$. The performance of DIOR-ViT and others on $P_{TestII}$, in general, improved upon the results on $P_{TestI}$. Similar observations were made in the previous study \citet{le2021joint}. Such discrepancy may be ascribable to the difference between the data quality and distributions. To identify the factors that are associated with this, a thorough investigation is needed. Additional independent tests may further provide insights into this issue.

In the analysis of DIOR-ViT with respect to different loss functions, the utility of the proposed $\mathcal{L}_{NAD}$ was apparent, outperforming all other combinations of loss functions. Moreover, we observed the effectiveness of multi-task learning in cancer classification. The models that are equipped with two (or three) loss functions, one for categorical classification and the others for differential ordinal classification, were generally superior to the model with a categorical classification loss function only. This is also aligned with the comparable performance by the previous multi-task learning approach ($\mathcal{M}_{MSE-CE_o}$ and $\mathcal{M}_{MSE-CE_o}$). Hence, further research and investigation of multi-task learning approaches may provide an additional performance gain in cancer grading.

\textcolor{black}{
Moreover, DIOR-ViT can be considered a variant of Siamese networks since it shares a feature extract to process two input samples separately and combines the resultant embedding vectors to produce the output (Here, the differential ordinal prediction). 
Siamese networks have been successfully applied to various applications such as visual tracking, image retrieval and matching, and pose estimation \citet{li2022survey} \citet{chicco2021siamese}. The typical Siamese network learns the similarity between the two input samples \citet{nandy2020survey}. The results of DIOR-ViT demonstrates that Siamese networks can be extended to learn the differential relationships among input samples. 
The differential ordinal learning of DIOR-ViT requires the pair-wise comparisons among input samples that can increase the complexity of the model and pose challenges for the model training. However, the comparison of the model complexity between DIOR-ViT and others (Table \ref{table:model_complexity}) demonstrates that the addition of the differential ordinal learning does not result in a substantial increment in the complexity. 
Nevertheless, DIOR-ViT can benefit from a more efficient training strategy, especially as it handles a larger dataset such as WSIs. We will further investigate this in the follow-up study.
}

\textcolor{black}{
We investigated the effectiveness of DIOR-ViT on three types of patch image datasets that were originally obtained from TMAs and WSIs. Recently, there have been great interests in the WSI-level classification. For example, \citet{wang2019rmdl} proposed RDML for gastric cancer classification. It utilizes a localization network to select the discriminative patches and then the patch features are aggregated by local-global feature fusion, recalibration module, and multi-instance pooling for the WSI-level classification.
\citet{lu2021data} developed CLAM for the WSI-level classification. CLAM uses attention mechanism to identify important regions and clustering to refine the feature space. 
These methods generally adopt multiple instance learning to process WSIs. 
DIOR-ViT should be applicable to various WSI-level classification tasks as many of such tasks involve the ordered class labels such as normal, dysplasia, and cancer in the gastric cancer classification \citet{wang2019rmdl}. However, this is beyond the scope of this study. We leave this for the future study.
}

\begin{table}[]
\centering
\caption{Comparison of model complexity metrics.}
\resizebox{\textwidth}{!}{
\begin{tabular}{cccccc}
    \hline
    \textbf{Model} & \textbf{\#Params(M)} & \textbf{\#Training FLOPs(M)} & \textbf{Training (ms/image)} & \textbf{\#Inference FLOPs(M)} & \textbf{Inference (ms/image)} \\ \hline
    ResNet & 24.7 & 12,175.96 & 88.11 & 12,077.97 & 38.07 \\
    DenseNet & 7.6 & 8,556.98 & 151.52 & 8,418.85 & 62.81 \\
    EfficientNet & 4.8 & 1,228.82 & 106.72 & 1,169.42 & 38.71 \\
    MSBP-Net & 24.8 & 4,583.72 & 89.13 & 4,550.38 & 41.08 \\
    $\mathcal{M}_{MSE-CE_o}$ & 4.8 & 1,228.82 & 108.23 & 1,169.42 & 46.15 \\
    $\mathcal{M}_{MAE-CE_o}$ & 4.8 & 1,228.82 & 109.41 & 1,169.42 & 45.70 \\
    ViT & 98.7 & 67,425.72 & 126.90 & 67,425.72 & 56.50 \\
    Swin & 87.5 & 47,191.15 & 125.31 & 47,191.15 & 44.42 \\
    Deit III & 304.0 & 191,210.69 & 138.89 & 191,210.69 & 48.95 \\
    GLRGC & 21.7 & 10,820.47 & 79.94 & 10,787.51 & 40.49 \\
    StoHisNet & 31.1 & 7,816.22 & 78.31 & 7,786.95 & 61.96 \\
    DIOR-ViT (Ours) & 98.7 & 67,425.72 & 131.41 & 67,425.72 & 56.53 \\
    DIOR-Swin (Ours) & 87.5 & 47,191.15 & 126.58 & 47,191.15 & 44.58 \\ \hline
\end{tabular}%
}
\label{table:model_complexity}
\end{table}

\section{Conclusions}
\textcolor{black}{
In this work, we propose a transformer-based neural network, DIOR-ViT, that simultaneously conducts both categorical classification and differential ordinal classification for cancer grading in computational pathology. Introducing differential ordinal learning, DIOR-ViT bridges the gap between the conventional categorical classification and the practice of cancer grading in pathology. The experimental results on three types of cancer datasets confirm the validity of DIOR-ViT and differential ordinal learning, which is not only able to identify cancer tissues of differing grades with high accuracy but also able to adapt well to unseen data. These results also demonstrate the importance of pathology knowledge and its integration into computational pathology tools for improved diagnosis. Cancer grading is not the only problem that can be formulated as a differential ordinal learning problem. Other problems, diseases, and datasets in computational pathology and medical imaging in general can benefit from it as the class labels possess ordering relationships. The future study will entail the further development and application of the proposed method to other problems and domains.
}
\label{sec:methods}

\section*{Acknowledgment}
This work was supported by a grant of the National Research Foundation of Korea (NRF) (No. 021R1A2C2014557, No. 2021R1A4A1031864, and No. RS-2024-00397293) and the Korea Health Technology R\&D Project through the Korea Health Industry Development Institute (KHIDI), funded by the Ministry of Health \& Welfare, Republic of Korea (No. HI21C1137).

\bibliographystyle{apalike}
\bibliography{refs}

\end{document}